\documentclass{PoS}
\def\d{\mathrm{d}}
\usepackage{slashed}
\usepackage{subfigure}
\def\d{\mathrm{d}}
\def\beq{\begin{equation}}
\def\eeq{\end{equation}}
\def\bea{\begin{eqnarray}}
\def\eea{\end{eqnarray}}

\def\nnb{\nonumber}
\def\ga{\left(}
\def\dr{\right)}

\def\nnb{\nonumber}

\def\ba{\begin{array}}
	\def\ea{\end{array}} 
 
\title{AdS/QCD predictions for semileptonic and rare B decays to $\rho$ and $K^*$ vector mesons } 
 
\ShortTitle{AdS/QCD predictions for semileptonic and rare B decays to $\rho$ and $K^*$ vector mesons } 
 
\author{\speaker{Mohammad Ahmady} 
         \\ 
        Department of Physics, Mount Allison University, Sackville, New Brunswick, Canada E4L 1E6\\ 
        E-mail: \email{mahmady@mta.ca}} 

\author{S\'{e}bastien Lord 
         \\ 
        D\'epartement de Math\'ematiques et Statistique, Universit\'e de Moncton,
        Moncton, N-B. E1A 3E9, Canada\\ 
        E-mail: \email{esl8420@umoncton.ca}} 
 
\author{Ruben Sandapen 
         \\ 
        Department of Physics, Mount Allison University, Sackville, New Brunswick, Canada E4L 1E6\\ 
        E-mail: \email{rsandapen@mta.ca}} 

\abstract{The light front wavefunction obtained from anti-de Sitter Quantum Chromodynamics is used to calculate the light cone distribution amplitudes for $\rho$ and $K^*$ vector mesons.  These distribution amplitudes are then utilized to calculate the $B\to\rho ,\; K^*$ transition form factors via light cone sum rules.   Two-parameter fits of our results for low to intermediate momentum transfer $q^2$ that includes the QCD lattice data at high $q^2$ are presented.  Consequently, we give predictions for the semileptoic $B\to\rho\ell\nu$ and dileptonic $B\to K^*\mu^+\mu^-$ decays. } 
 
\FullConference{XXIII International Workshop on Deep-Inelastic Scattering\\ 
		 27 April - May 1 2015\\ 
		 Dallas, Texas} 
 
\begin{document} 
 
\section{Introduction} 
Rare B meson decays have been at the focus of a significant number of theoretical and experimental investigations. The underlying quark processes for these decays are suppressed either due to small Cabbibo-Kobayashi-Mashawa (CKM) quark mixing or lack of flavor changing neutral current (FCNC) at tree level in standard model (SM).  A better theoretical understanding of the former decays can help with our knowledge of the least known CKM matrix elements like $V_{ub}$.  On the other hand, since the latter processes occur only through quantum loops, they are sensitive to new unknown particles that are predicted by various new physics (NP) scenarios beyond the SM.  These so-called exotic particles can appear as virtual entries in the loops and alter the decay rate and other associated observables from what is expected within the SM.  This indirect search for NP in rare B meson decays has been used to constrain various scenarios like supersymmetry and vector-like quark model.

Here, we report a number of predictions on semileptonic and rare dileptonic B decays to light vector mesons based on anti-de Sitter Quantum Chromodynamics (AdS/QCD)\cite{PRD3,PRD4,PRD5}.  The holographic light-front wavefunctions obtained from AdS/QCD are used to calculate the distribution amplitudes (DAs) of the $\rho$ and $K^*$ vector mesons.  These DAs are then inserted in light-cone sum rules (LCSR) formulas for $B\to \rho ,\; K^*$ transition form factors.  Our results contain predictions for a number of observables related to $B\to \rho\ell\nu$ and $B\to K^*\mu^+\mu^-$.

\section{Effective hamiltonian}
The effective Hamiltonian for $b\to u \ell\nu$, $b\to s\gamma$ and $b\to s\ell\bar{\ell}$ transitions(Replace $s$ with $d$ for $b\to d(\gamma ,\; \ell\bar{\ell})$ transition) can be written as:
\begin{equation}\label{heff}
{\cal H}_{eff}=\frac{G_F}{\sqrt{2}}\sum_{p=u,c}\left\{ V^*_{ps}V_{pb}
\left[ C_1 Q^p_1 + C_2 Q^p_2 +\sum_{i=3,\ldots ,10} C_i Q_i\right]+V_{pb}Q^p_0\right\}
\end{equation}
where the operators $Q_i,\; i=0..10$ are defined as the following:

\begin{eqnarray}\label{q1def}
    Q^p_0 &=& (\bar \nu\ell)_{V-A}(\bar pb)_{V-A}\nonumber \\
	Q^p_1 &=& (\bar sp)_{V-A}(\bar pb)_{V-A} \;\;\;\;\;\;\;\;\;\;\;
	Q^p_2 = (\bar s_i p_j)_{V-A}(\bar p_j b_i)_{V-A} \nonumber\\
	Q_3 &=& (\bar sb)_{V-A} \sum_q (\bar qq)_{V-A} \;\;\;\;\;\;
	Q_4 =   (\bar s_i b_j)_{V-A} \sum_q (\bar q_j q_i)_{V-A} \nonumber\\
	Q_5 &=& (\bar sb)_{V-A} \sum_q (\bar qq)_{V+A} \;\;\;\;\;\;
	Q_6 = (\bar s_i b_j)_{V-A} \sum_q (\bar q_j q_i)_{V+A} \nonumber\\
	\label{q7def}
	Q_7 &=& \frac{e}{8\pi^2}m_b\,
	\bar s_i\sigma^{\mu\nu}(1+\gamma_5)b_i\, F_{\mu\nu}\;\;
	\label{q8def}
	Q_8 = \frac{g}{8\pi^2}m_b\,
	\bar s_i\sigma^{\mu\nu}(1+\gamma_5)T^a_{ij} b_j\, G^a_{\mu\nu}\nonumber \\
	Q_9 &=& \frac{e^2}{8\pi^2}
	\bar s\gamma^{\mu}(1-\gamma_5)b\, \bar{\ell}\gamma^\mu\ell\;\;\;\;\;\;
	\label{q8def}
	Q_{10} = \frac{e^2}{8\pi^2}
	\bar s\gamma^{\mu}(1-\gamma_5)b\, \bar{\ell}\gamma^\mu\gamma_5\ell
\end{eqnarray}
$Q^p_0$ is relevant to semileptonic decay and $Q_i,\; i=1..10$ appear in rare radiative and dileptonic decays.  The effective Hamiltonian \ref{heff} is obtained by integrating out the particles much heavier than the B meson mass, i.e. $W$ and $Z$ bosons as well as the top quark.  The Wilson coefficients $C_i$ are evaluated perturbatively at next-to-leading order (NLO) and their numerical values are shown in Table \ref{tab:Wilson_coef}.

\begin{table}
	\begin{tabular}{| c | c | c | c | c | c | c | c | c | c | c |}
		\hline
		$C_{1}$ & $C_{2}$ & $C_{3}$ & $C_{4}$ & $C_{5}$ & $C_{6}$ & $C_{7}^{eff}$ & $C_{8}^{eff}$ & $C_{9}$& $C_{10}$\\
		\hline \hline
		$-0.148$ & $1.060$ & $0.012$ & $-0.035$ & $0.010$ & $-0.039$ & $-0.307$ & $-0.169$ & $4.238$ & $-4.641$ \\ \hline
		
	\end{tabular}
	\caption{Tabulated Wilson coefficeints.}
	\label{tab:Wilson_coef}
\end{table}
The nonperturbative QCD effects appear in the $B\to V,\; V=\rho ,\; K^*$ transition matrix elements which are parametrized in terms of a number of form factors.  Understanding these form factors is crucial in theoretical calculation of the exclusive B meson decays to light mesons. 

\section{$B\to \rho ,\; K^*$ transition form factors}
From the structure of the operators in \ref{q1def} we observe that the transition matrix element is either of $V-A$ or tensor forms.  These two forms can be parametrized in terms of 7 form factors as the following: 
	 
		\begin{eqnarray}
		\langle V (k,\varepsilon)|\bar{q} \gamma^\mu(1-\gamma^5 )b | B(p) \rangle &=& \frac{2i V(q^2)}{m_B + m_{K^*}} \epsilon^{\mu \nu \rho \sigma} \varepsilon^*_{\nu} k_{\rho} p_{\sigma} -2m_{K^*} A_0(q^2) \frac{\varepsilon^* \cdot q}{q^2} q^{\mu}  \nonumber \\
		&-& (m_B + m_{K^*}) A_1(q^2) \left(\varepsilon^{\mu *}- \frac{\varepsilon^* \cdot q q^{\mu}}{q^2} \right) \nonumber \\
		&+& A_2(q^2) \frac{\varepsilon^* \cdot q}{m_B + m_{K^*}}  \left[ (p+k)^{\mu} - \frac{m_B^2 - m_{K^*}^2}{q^2} q^{\mu} \right] 
		\nonumber
		\end{eqnarray}
		\begin{eqnarray}
		q_{\nu} \langle V (k,\varepsilon)|\bar{q} \sigma^{\mu \nu} (1-\gamma^5 )b | B(p) \rangle &=& 2 T_1(q^2) \epsilon^{\mu \nu \rho \sigma} \varepsilon^*_{\nu} p_{\rho} k_{\sigma} \nonumber \\
		&-& i T_2(q^2)[(\varepsilon^* \cdot q)(p+k)_{\mu}-\varepsilon_{\mu}^*(m_B^2-m_{K^*}^2)] \nonumber \\
		&-& iT_3(q^2) (\varepsilon^* \cdot q) \left[ \frac{q^2}{m_B^2-m_{K^*}^2} (p+k)_{\mu} -q_{\mu}  \right] 
		\label{formfactors}
		\end{eqnarray}
	We use LCSR\cite{ali,lcsr} to calculate the seven form factors in terms of the DAs of the light vector mesons. LCSR are variations of the traditional QCD Sum Rules whereby non-local matrix elements are expanded in terms of light front DAs. For example, the LCSR for the radiative form factor $T_1$ is given below:
	\begin{eqnarray}
	T_1(q^2) &=& \frac{1}{4} \left( \frac{m_b}{f_B m_B^2}\right) \exp{\left(\frac{m_B^2}{M^2}\right)}
	\int_\delta^1 \frac{\d u}{u}\, \nnb \exp \ga - \frac{m_b^2 + p^2 u \bar u - q^2 \bar u}{uM^2} \dr \Bigg\{ m_b f_V^\perp \phi_\perp (u) + \nnb \\&&
	\; f_V m_V \Bigg[ \Phi_\parallel (u)  +
	u g_\perp^{(v)} (u)  +\frac{g_\perp^{(a)}(u)}{4}  + \frac{(m_b^2 + q^2 -p^2 u^2)g_\perp^{(a)} (u)}{4 u M^2} \Bigg] \Bigg\}
	\label{LCSRT1}
	\end{eqnarray}
	In Eqn. \ref{LCSRT1}, $M$ is the Borel parameter and $\delta$ is associated with the continuum threshold \cite{ali}.  Using light-cone coordinates, $x^{\pm}=x^0\pm x^3$, $x^\perp =x^1,\; x^2$, the two twist-2 DAs $\phi_{\perp , \parallel}$ along with the two twist-3 DAs $g_\perp ^{(v,a)}$ are defined through the following relation\cite{Ball:2007zt}:
\begin{eqnarray}
\langle 0|\bar q(0)  \gamma^\mu q(x^-)| V
(P,\lambda)\rangle \nonumber 
&&= f_{V} M_{V}
\frac{e_{\lambda} \cdot x}{P^+x^-}\, P^\mu
\int_0^1 \mathrm{d} u \; e^{-iu P^+x^-}
\phi_\parallel(u,\mu)
 \\
&&+ f_{V} M_{V}
\left(e_{\lambda}^\mu-P^\mu\frac{e_{\lambda} \cdot
	x}{P^+x^-}\right)
\int_0^1 \mathrm{d} u \; e^{-iu P^+x^-} g_\perp^{ (v)}(u,\mu) \;,
\label{DA:phiparallel-gvperp}
\end{eqnarray}
\begin{eqnarray}
\langle 0|\bar q(0) [\gamma^\mu,\gamma^\nu] q (x^-)|V
(P,\lambda)\rangle  
=2 f_{V}^{\perp} (e^{\mu}_{\lambda} P^{\nu} -
e^{\nu}_{\lambda} P^{\mu}) \int_0^1 \mathrm{d} u \; e^{-iuP^+ x^-} \phi_{\perp}
(u, \mu)  \;,
\label{DA:phiperp}
\end{eqnarray}
\begin{eqnarray}
\langle 0|\bar q(0) \gamma^\mu \gamma^5 s(x^-)|V (P,\lambda)\rangle 
=-\frac{1}{4} \epsilon^{\mu}_{\nu\rho\sigma} e_{\lambda}^{\nu}
P^{\rho} x^{\sigma}  \tilde{f}_{V} M_{V} \int_0^1 \mathrm{d} u \; e^{-iuP^+ x^-}
g_\perp {(a)}(u, \mu)  \;,
 \label{DA:gaperp}
\end{eqnarray}
where 
\begin{equation}
\tilde{f}_{\rho} = f_{\rho}\;,
\end{equation}
and
\begin{equation}
\tilde{f}_{K^*} = f_{K^*}-f_{K^*}^{\perp} \left(\frac{m_s + m_{\bar{q}}}{M_{K^*}} \right) \;.
\end{equation}	
$m_s$ and $m_{\bar q}$ are the masses of the strange quark and light anti-quark, respectively.  As $x^-\to 0$, in Eqns \ref{DA:phiparallel-gvperp} and \ref{DA:phiperp} we recover the usual definition for the decay constant $f_V$ and $f^\perp_V$:
\begin{equation}
\langle 0|\bar q(0)  \gamma^\mu q(0)|V
(P,\epsilon)\rangle
= f_\rho M_V \epsilon^\mu \; ,
\label{decayconstant}
\end{equation}
\begin{equation}
\langle 0|\bar q(0) [\gamma^\mu,\gamma^\nu] q(0)|V (P,\epsilon)\rangle =2 f_V^{\perp} (\epsilon^{\mu} P^{\nu} - \epsilon^{\nu} P^{\mu})\; .
\label{tensordecayconstant}
\end{equation}
Also, all the above-defined DAs are normalized, i.e.
$$
\int_0^1 \d u \;
\phi_{\rho}^{\perp ,\parallel}(u,\mu)=	\int_0^1 \d u \; g^{\perp (a,v)}_{\rho}(u, \mu)=1
$$
The DAs which are commonly used in the literature are derived from QCD sum rules (SR).  In the next section, we put forward alternative DAs which are obtained from AdS/QCD.

\section{AdS/QCD predictions for $\rho$ and $K^*$ DAs}
The holographic light-front wavefunction for a vector meson $(L=0,S=1)$ in AdS/QCD can be written as\cite{Brodsky}:
\begin{eqnarray}
\phi_{\lambda} (z,\zeta) \propto \sqrt{z(1-z)} \exp
\left(-\frac{\kappa^2 \zeta^2}{2}\right)
\times\exp\left \{-\left[\frac{m_q^2-z(m_q^2-m^2_{\bar{q}})}{2\kappa^2 z (1-z)} \right] 
\right \} \label{AdS-QCD-wfn}
\end{eqnarray}
with $\kappa=M_{V}/\sqrt{2}$ and $m_q=m_{\bar q},\; m_s$ for $\rho$ and $K^*$, respectively.  $\lambda$ is the polarization of the vector meson.  We have introduced the dependence on quark masses following a prescription by Brodsky and de T\'eramond \cite{Brodsky2}.  This wavefunction for the $\rho$ meson was successfully used to predict the diffractive $\rho$ meson electroproduction at HERA\cite{PRL}.  

\begin{figure}
	\centering
	\subfigure[~Twist-$2$ DA for the longitudinally polarized $\rho$ meson]{\includegraphics[width=.350\textwidth]{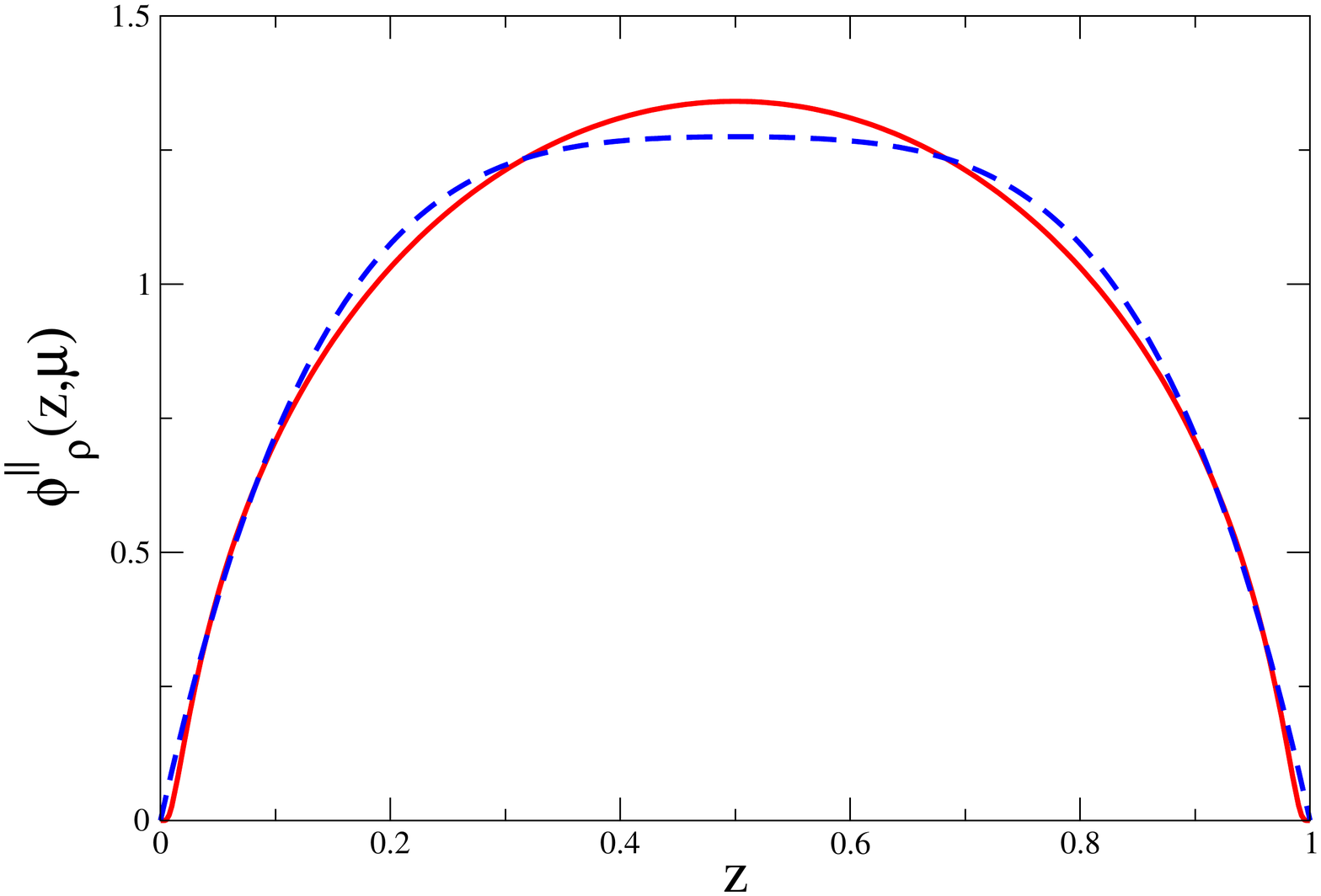} }
	\subfigure[~Twist-$2$ DA for the transversely polarized $\rho$ meson]{\includegraphics[width=.350\textwidth]{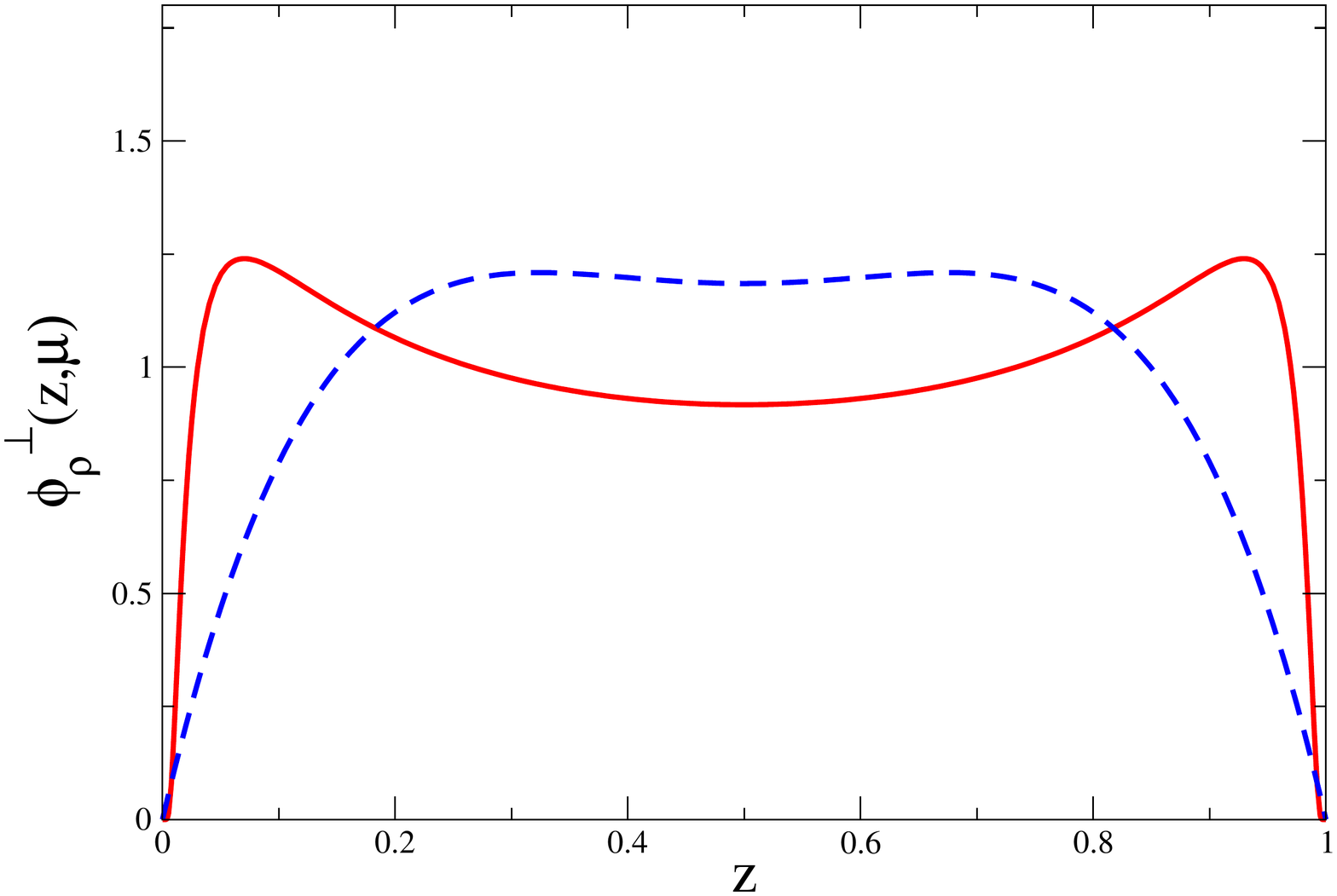} }
	\caption{Twist-$2$ DAs for the $\rho$ meson. Solid: AdS/QCD DAs; Dashed: Sum Rules DAs.} \label{fig:tw2DAs}
\end{figure}

\begin{figure}
	\centering
	\subfigure[~Twist-$2$ DA for the longitudinally polarized $K^*$ meson]{\includegraphics[width=.350\textwidth]{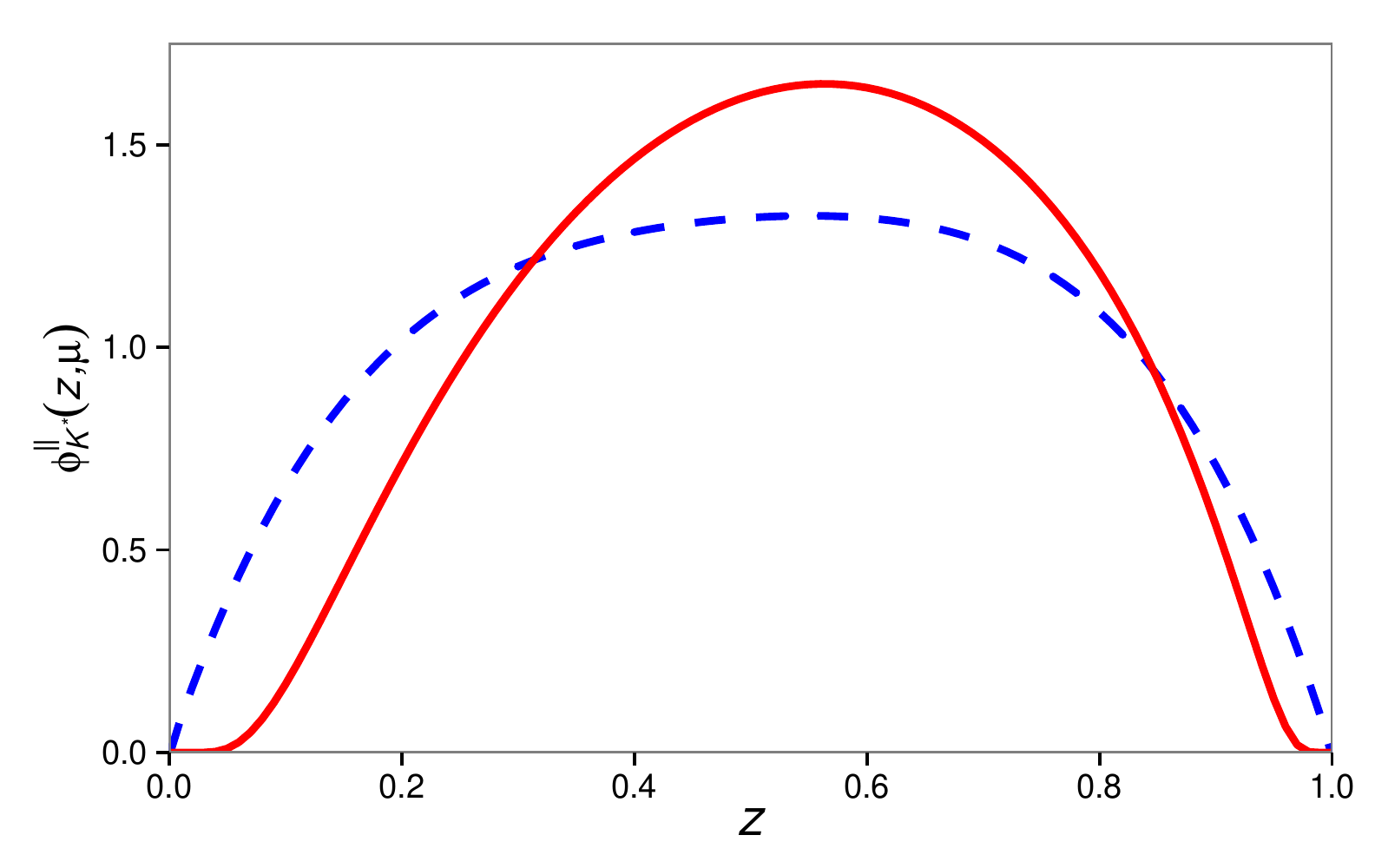} }
	\subfigure[~Twist-$2$ DA for the transversely polarized $K^*$ meson]{\includegraphics[width=.350\textwidth]{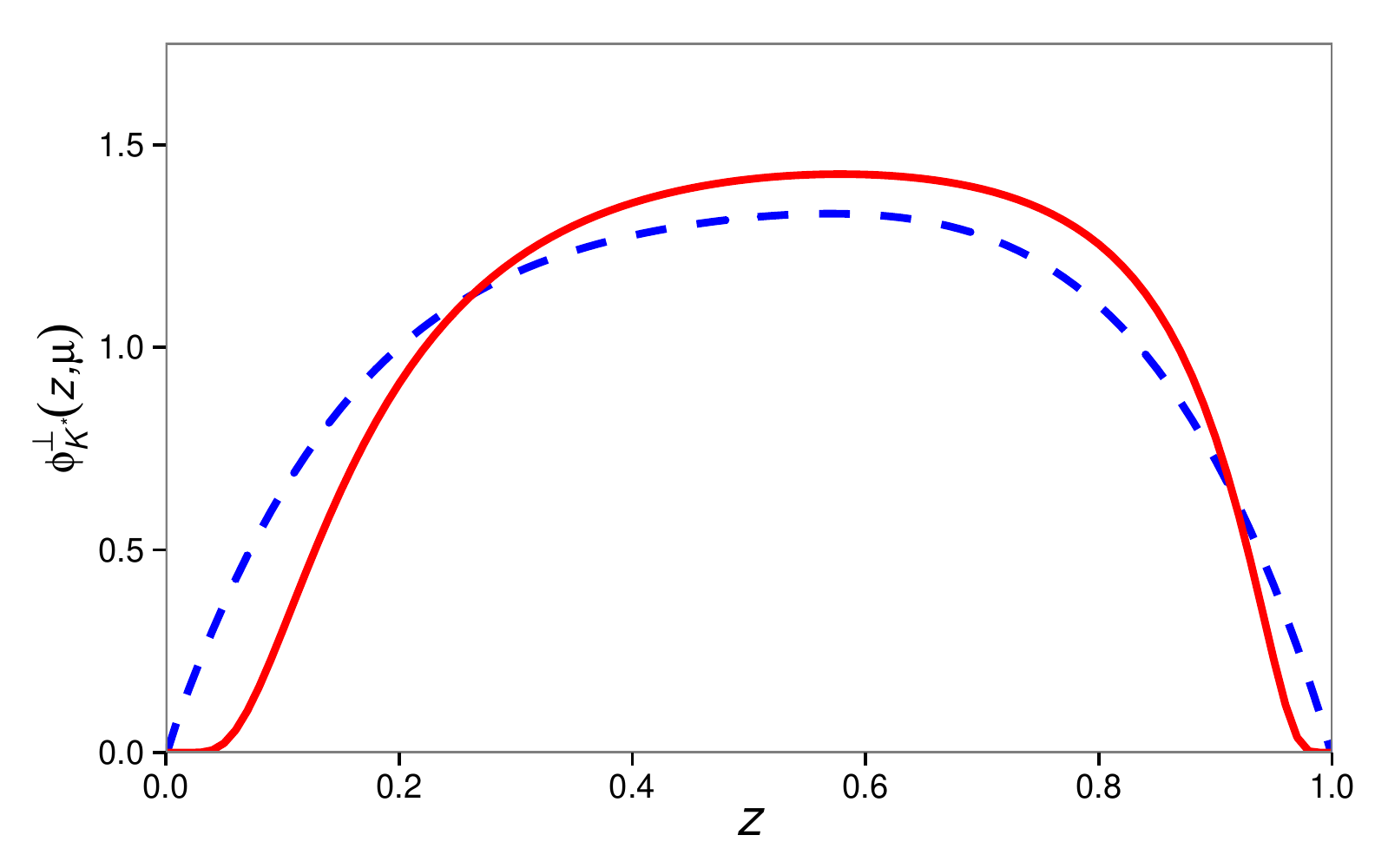} }
	\caption{Twist-$2$ DAs for the $K^*$ meson. Solid: AdS/QCD DAs; Dashed: Sum Rules DAs.} \label{fig:tw2DAsKstar}
\end{figure}

The relation between the wavefunction \ref{AdS-QCD-wfn} and DAs for $\rho$ and $K^*$ was derived in \cite{PRD1,PRD2}.  Figures \ref{fig:tw2DAs} and \ref{fig:tw2DAsKstar} compare the two twist-2 DAs for these two vector mesons obtained from AdS/QCD and SR.

\begin{figure}
	\centering
	{\includegraphics[width=.40\textwidth]{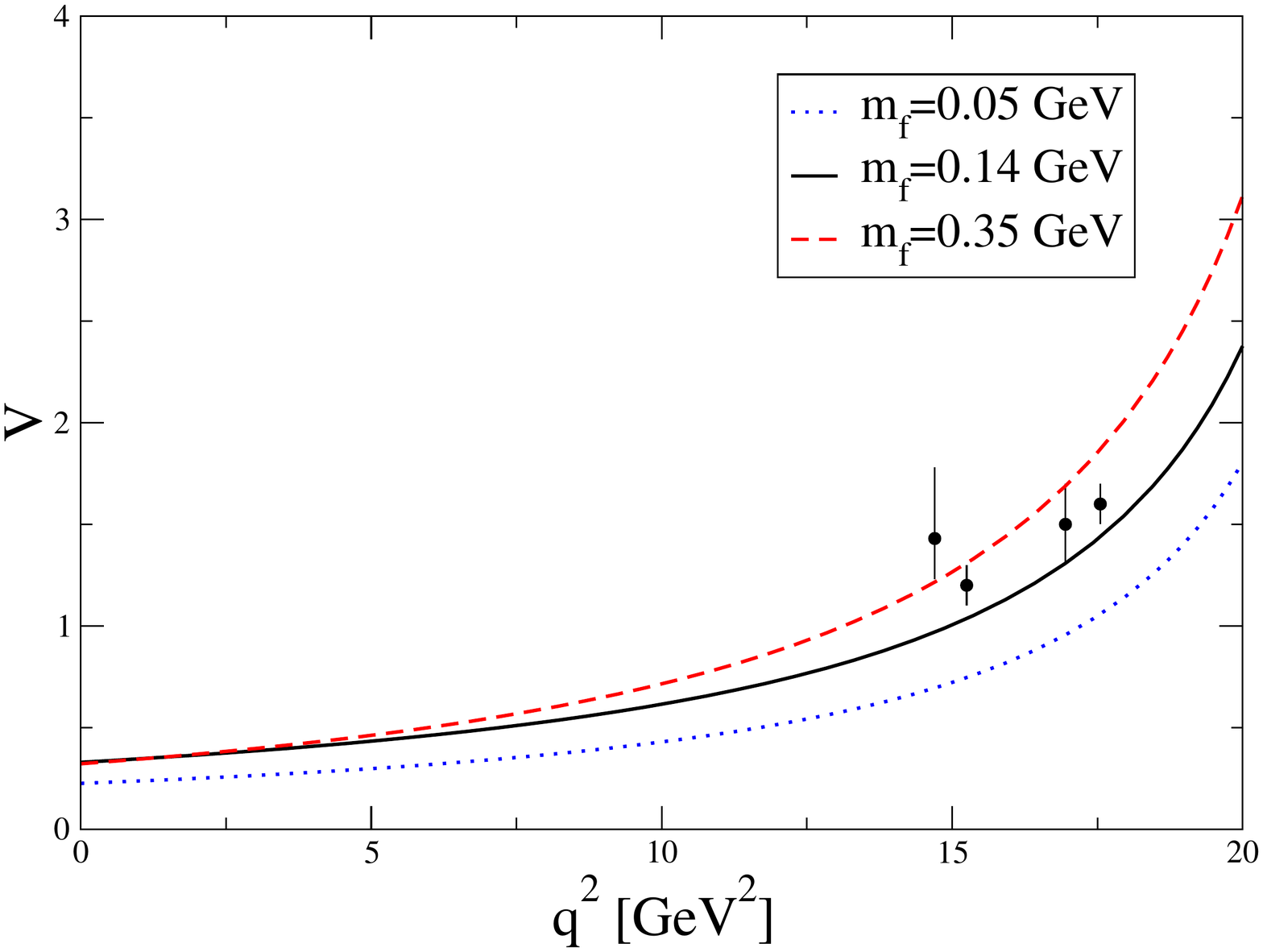} }
	{\includegraphics[width=.40\textwidth]{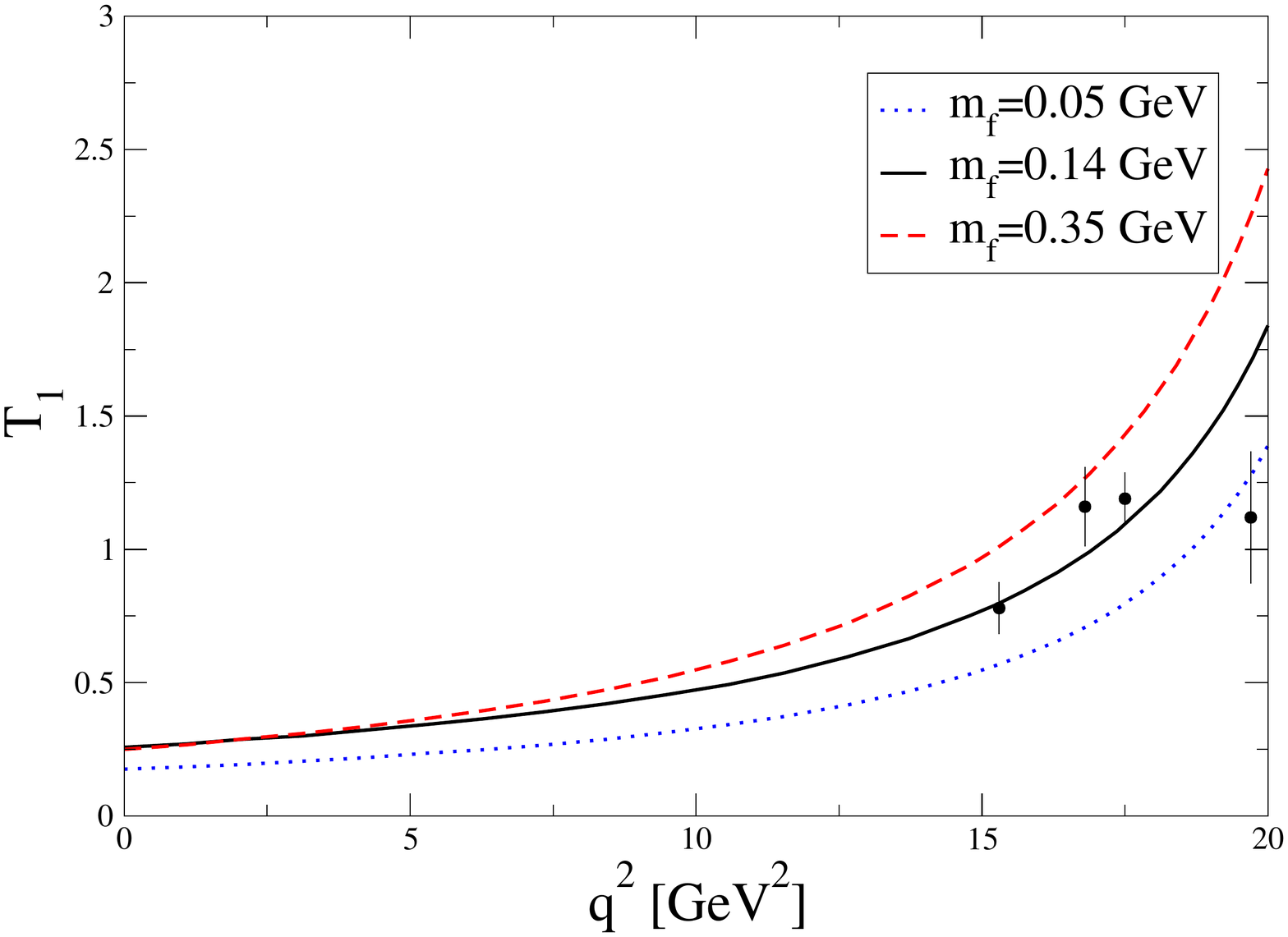} }
	\caption{The AdS/QCD prediction for $B\to\rho$ transition form factors $V$ and $T_1$ for 3 different quark mass inputs.  The available lattice data at high $q^2$ are shown as well.  } \label{formfactorsrho}
\end{figure}

Consequently, the AdS/QCD predictions for $B\to\rho ,\; K^*$ form factors can be computed via LCSR.  One should note that LCSR results are valid at low to intermediate values of the momentum transfer $q^2$.  Figure \ref{formfactorsrho} illustrates the AdS/QCD predictions for two $B\to\rho$ form factors $V$ and $T_1$ when 3 different values of the quark mass are used.  Our results for the all 7 form factors for this transition can be found in \cite{PRD3}.  The data points on this figure are from lattice calculation which are available at high $q^2$\cite{Lattice}.  For our numerical calculations of the semileptonic $B\to\rho\ell\nu$ decay, we find two-parameter fits for the form factors using AdS/QCD predictions at low to intermediate $q^2$ and lattice data at high $q^2$\cite{PRD3}. 

\begin{figure}
	\centering
	{\includegraphics[width=.40\textwidth]{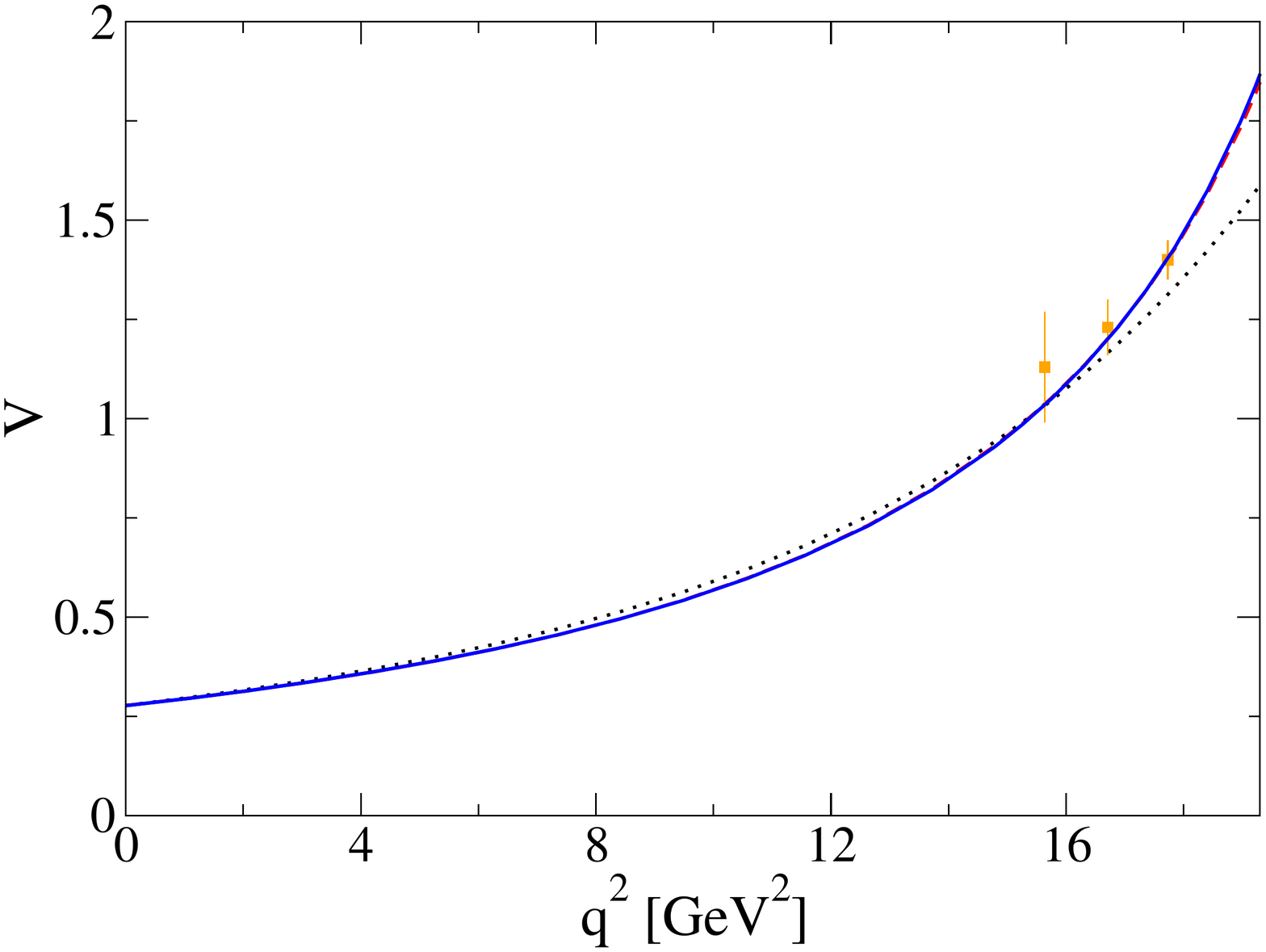} }
	{\includegraphics[width=.40\textwidth]{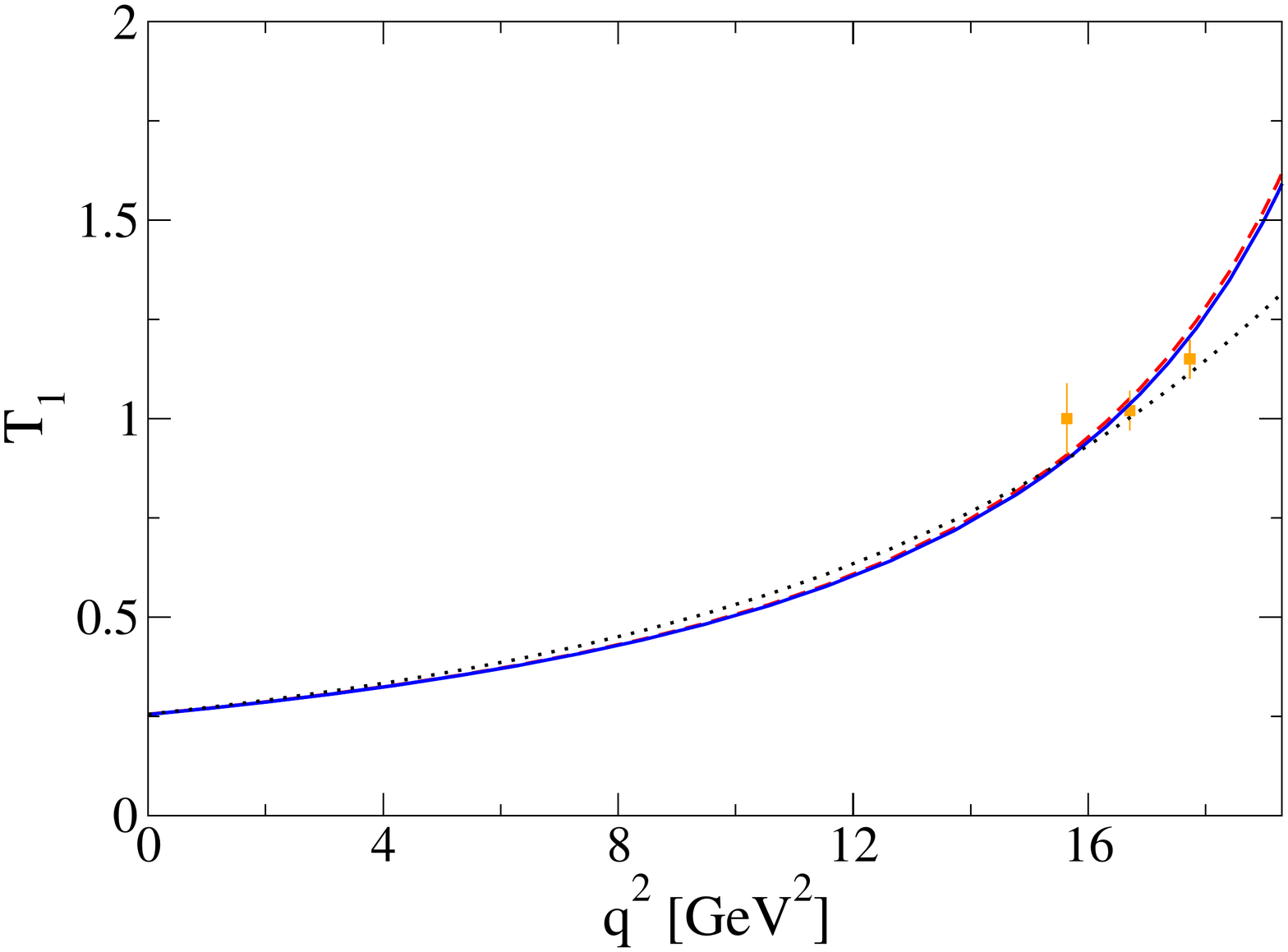} }
	\caption{The AdS/QCD  prediction for $B\to K^*$ transition form factors $V$ and $T_1$. The solid curve denotes AdS/QCD. The dashed curve denotes the AdS/QCD fit. The dotted curve denotes the fit to AdS/QCD and lattice. The available lattice data at high $q^2$ are shown as well. } \label{formfactorsKstar}
\end{figure}

Figure \ref{formfactorsKstar} shows the AdS/QCD prediction for $B\to K^*$ form factors $V$ and $T_1$ as compared with those obtained from SR.  The data points at high $q^2$ are from lattice calculations\cite{latticekstar}.  Our results for the full set of $B\to K^*$ transition form factors can be found in \cite{PRD4}.  Similar to $B\to\rho$ case, we use two-parameter fits to AdS/QCD and lattice data combined for the form factors inserted in our numerical computations.
 
\section{Numerical predictions}

The differential decay rate of the semileptonic $B\to \rho\ell\nu$ is sensitive to $V_{ub}$.  
BaBar collaboration has measured the partial branching fractions for this decay channel in three $q^2$ bins\cite{babar}: $\Delta B_{\mbox{\tiny{low}}}=(0.564\pm 0.166)\times 10^{-4},\; \Delta B_{\mbox{\tiny{mid}}}=(0.912\pm 0.147)\times 10^{-4}$ and $\Delta B_{\mbox{\tiny{high}}}=(0.268\pm 0.062)\times 10^{-4}$ for $0<q^2<8,\; 8<q^2<16$ and $16<q^2<20.3\; {\rm GeV}^2$ respectively.  To eliminate the uncertainty in $V_{ub}$ in comparing our results with the above data, we take the ratios of the partial branching fractions as defined below:
\begin{equation}
R_{\mbox{\tiny{low}}}=\frac{\Delta B_{\mbox{\tiny{low}}}}{\Delta B_{\mbox{\tiny{mid}}}}=0.618\pm 0.207 ,\;\;\; R_{\mbox{\tiny{high}}}=\frac{\Delta B_{\mbox{\tiny{high}}}}{\Delta B_{\mbox{\tiny{mid}}}}=0.294 \pm 0.083
\end{equation}
AdS/QCD predictions for these ratios are $R_{\mbox{\tiny{low}}}=0.580, 0.424$ and $R_{\mbox{\tiny{high}}}=0.427,0.503$ for $m_q=0.14$ GeV and  $0.35$ GeV, respectively.  It seems that better agreement with data is achieved at low $q^2$.

\begin{figure}
	\centering
	{\includegraphics[width=.40\textwidth]{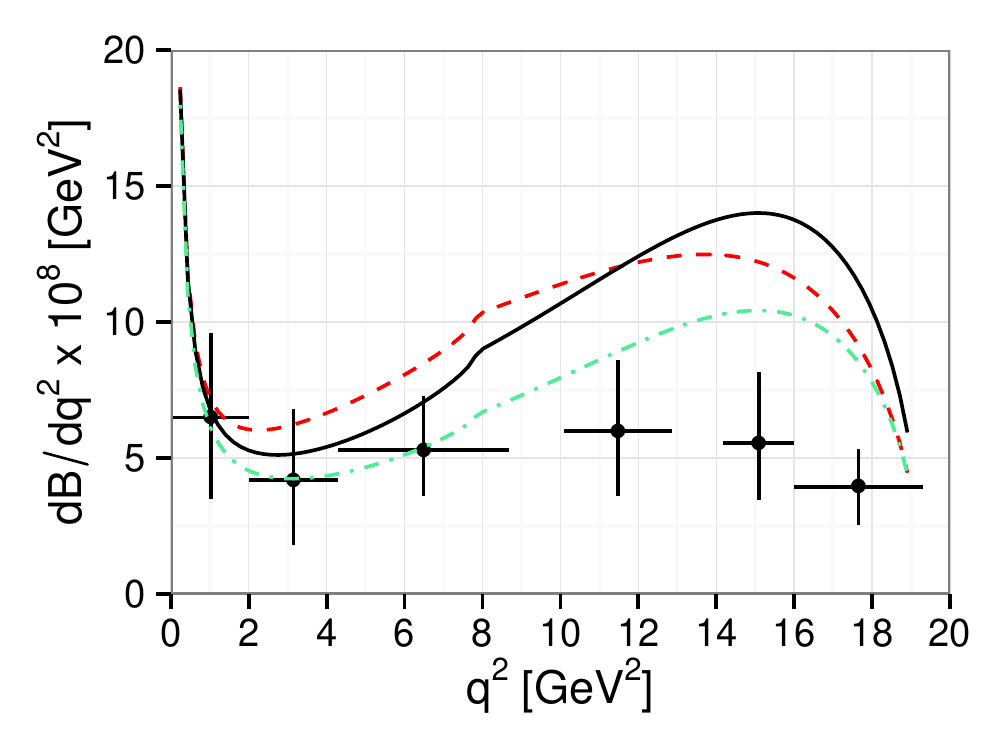} }
	{\includegraphics[width=.450\textwidth]{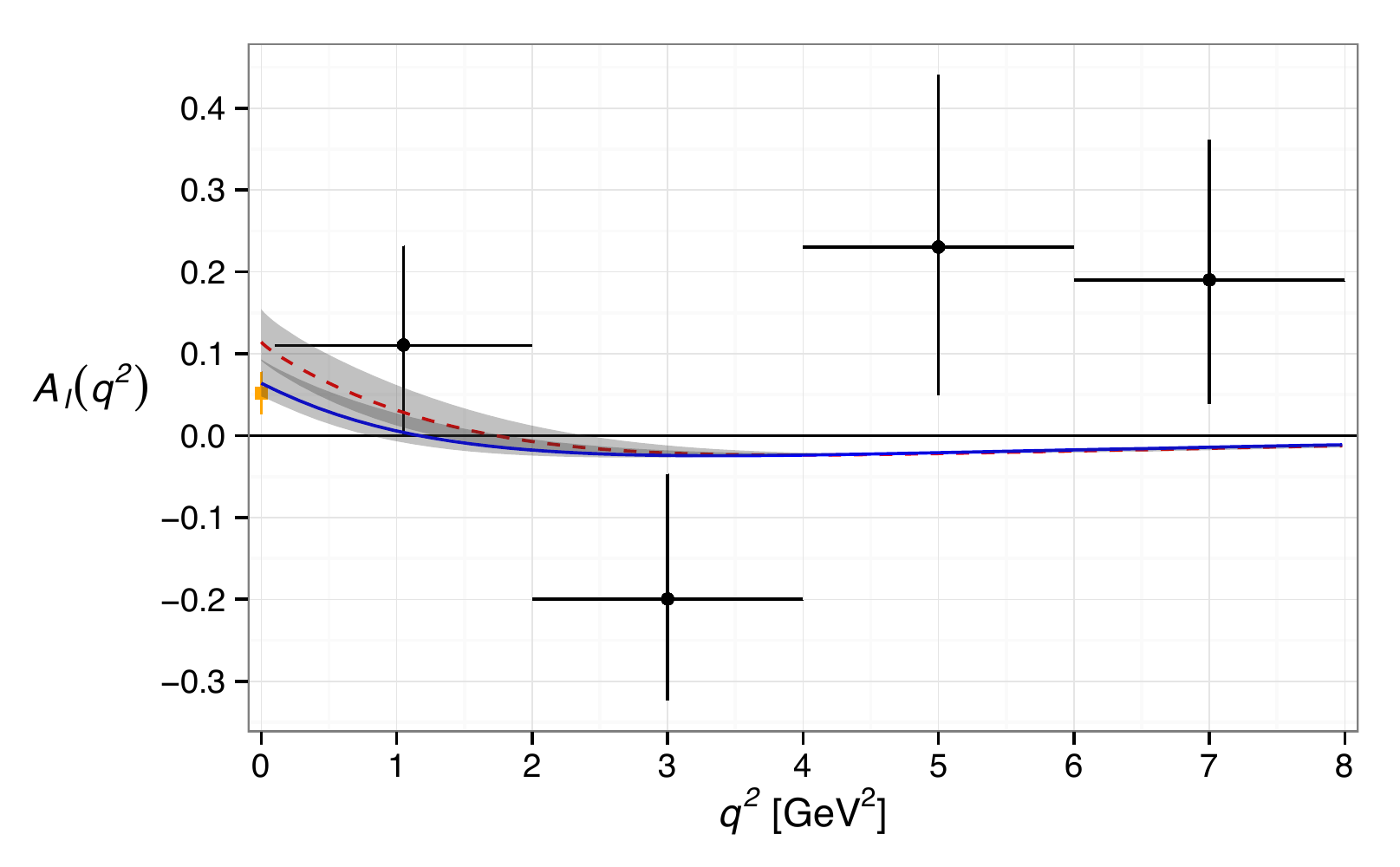} }
	\caption{The AdS/QCD  prediction for $B\to K^*\mu^+\mu^-$ differential decay width and isospin asymmetry distribution.  Differential decay rates are predicted with (solid curve) and without (dashed curve) lattice data input in two-parameter fits of the form factors as well as $C_9^{\rm NP}=-1.5$ (dash-dotted curve).  In the Isospin asymmetry graph, AdS/QCD prediction (solid curve) is presented with an uncertainty band and compared with SR expectation (dashed curve). } \label{BKstarmu}
\end{figure}

Our results for the differential decay rate and isospin asymmetry distribution in $B\to K^*\mu^+\mu^-$ is shown in Figure \ref{BKstarmu}.  Data points for the decay rate are from LHCb\cite{LHCb} and the predictions are given with (solid curve) and without (dashed curve) considering lattice data in two-point fits of the form factors\cite{PRD4}. We also present our prediction for this observable assuming NP contribution to the Wilson coefficient $C_9$ (dash-dotted curve).  The AdS/QCD prediction for isospin asymmetry distribution in $B\to K^*\mu^+\mu^-$ in Figure \ref{BKstarmu} is presented (solid curve) with an uncertainty band due to the renormalization scale dependence. The data points are from Belle and BABAR\cite{bellebabar}.  For comparison, we also include the SR predictions (dashed curve) for isospin asymmetry distribution.  Our prediction for asymmetry at $q^2=0$ (relevant to $B\to K^*\gamma$) is consistent with experimental data.

\end{document}